
\input harvmac

\def\muo{\mu _{out}}
\def\mui{\mu _{in}}
\def\k{\kappa}
\def\({\left (}
\def\){\right )}
\def\xa{\xi _1}
\def\xb{\xi _2}
\def\xc{\xi _3}
\def\xd{\xi _4}
\def\do{\delta _{out}}
\def\di{\delta _{in}}

\def\ep{\epsilon}
\def\ko{ (1-{1\over 2\pi}\delta_{out} ) }
\def\gp{G'(\xi _3 )}
\def\gpp{ G''(\xi _3 ) }
\def\0h{\hat{0}}
\def\D{\Delta}
\def\ajou#1&#2(#3){\ \sl#1\bf#2\rm(19#3)}
\lref\dual{A. Shapere, S. Trivedi and F. Wilczek, {\it Mod. Phys. Lett.} {\bf
6A}, 2677 (1991).}
\lref\hhr{S. W. Hawking, G. T.  Horowitz, and S. F. Ross, \ajou Phys. Rev.
&D51 (95) 4302.}
\lref\ernst{F. J. Ernst, \ajou J. Math. Phys. &17 (76) 515.}
\lref\kw{W. Kinnersley and M. Walker, \ajou Phys. Rev. & D2 (70) 1359.}
\lref\tev{A.Ashtekar and T. Dray}
\lref\gibbons{G.W. Gibbons,
in {\sl Fields and geometry}, proceedings of
22nd Karpacz Winter School of Theoretical Physics: Fields and
Geometry, Karpacz, Poland, Feb 17 - Mar 1, 1986, ed. A. Jadczyk (World
Scientific, 1986).}
\lref\gs{D. Garfinkle and A. Strominger,
\ajou Phys. Lett. &256B (91) 146.}
\lref\ggs{D. Garfinkle, S. Giddings and A. Strominger}
\lref\ruth{A. Achucarro, R. Gregory and K. Kuijken, ``Abelian Higgs hair
for black holes", gr-qc/9505039.}
\lref\dgkt{H.F. Dowker, J.P. Gauntlett, D. A. Kastor, J. Traschen,
\ajou Phys. Rev. &D49 (94) 2909.}
\lref\dggh{H.F. Dowker, J.P. Gauntlett, S. B. Giddings and G. Horowitz,
\ajou Phys. Rev. &D50 (94) 2662.}
\lref\by{J.D.Brown and J.W. York, {\it Phys. Rev.} {\bf D47} (1993) 1407;
{\it Phys. Rev.} {\bf D47}  (1993) 1420.}
\lref\hh{S. W. Hawking and G. Horowitz, ``The gravitational hamiltonian,
action, entropy, and surface terms", gr-qc/9501014.}
\lref\dgktinprep{F.Dowker, J.Gauntlett, D.Kastor,J.Traschen, in preparation.}
\lref\teit{ M. Banados, C. Teitelboim, J. Zanelli, \ajou Phys. Rev. Lett.
&72 (94) 957; C. Teitelboim, \ajou Phys. Rev. &D51 (95) 4315.}
\lref\vilenkin{A. Vilenkin, {\it Phys.\ Rep. } {\bf121} (1985) 263.}
\lref\jpav{ J. Preskill and A. Vilenkin,
{\it Phys. Rev.} {\bf D47} (1993) 2324.}
\lref\jppriv{John Preskill, private communication.}
\lref\ryder{L.H. Ryder, {\sl Quantum Field Theory} (Cambridge University
  Press, 1985), pp.415-417.}
\lref\dggh{H.F. Dowker, J.P. Gauntlett, S. B. Giddings and G. Horowitz,
\ajou Phys. Rev. &D50 (94) 2662.}
\lref\hr{S. W. Hawking and S. F. Ross, ``Pair production of black holes
on cosmic strings", gr-qc/9506020.}
\lref\emp{R. Emparan, ``Pair creation of black holes joined by cosmic
strings", gr-qc/9506025.}
\lref\qhair{S. Coleman, J. Preskill and F. Wilczek, {\it Nucl. Phys.}
{\bf B378} (1992) 175.}


%
%
\Title{\vbox{\baselineskip12pt
\hbox{NSF-ITP-95-48}
\hbox{UCSBTH-95-13}
\hbox{UMHEP-420}
\hbox{gr-qc/9506041}}}
{\vbox{\centerline{Breaking Cosmic Strings without Monopoles}
       }}
{
\baselineskip=12pt
\centerline{Douglas M. Eardley,$^1$ Gary T. Horowitz,$^2$
 David A. Kastor,$^{3a}$
Jennie Traschen$^{3b}$}
\bigskip
\centerline{\sl $^1$Institute for Theoretical Physics}
\centerline{\sl University of California}
\centerline{\sl Santa Barbara, CA 93106}
\centerline{\it doug@itp.ucsb.edu}
\bigskip
\centerline{\sl $^2$Department of Physics}
\centerline{\sl University of California}
\centerline{\sl Santa Barbara, CA 93106}
\centerline{\it gary@cosmic.physics.ucsb.edu}
\bigskip
\centerline{\sl $^3$Department of Physics and Astronomy}
\centerline{\sl University of Massachusetts}
\centerline{\sl Amherst, MA 01003-4525}
\centerline{\it $^a$Internet: kastor@phast.umass.edu}
\centerline{\it $^b$Internet: lboo@phast.umass.edu}
\bigskip
\centerline{\bf Abstract}
\medskip
It is shown that topologically stable cosmic strings can, in fact,
appear to end or to break, even in theories without monopoles. This can occur
whenever the spatial topology of the universe is
nontrivial.  For the case of Abelian-Higgs strings, we describe the
gauge and scalar field configurations necessary
for a string to end on
a black hole.  We give a lower bound for the
rate at which a cosmic string will break via
black hole pair production, using an instanton calculation based on the
Euclidean C-metric.

\Date{6/95}
\newsec{Introduction}
In the absence of
singularities or monopoles,
local cosmic strings cannot end, and hence, must either be infinite
in extent or form closed loops.  It is the purpose of this letter, however, to
point out that, if the topology of space is nontrivial, then local cosmic
strings may appear to end.  In particular, a cosmic string may
disappear down the throat of a black hole.
Moreover, when topology changing processes are included (as
suggested by quantum gravity)
a cosmic string can appear to break.

In a functional integral approach
to quantum gravity, the leading approximation to such a topology changing
process is given by an instanton, or solution to the Euclidean field equations,
which interpolates between the initial and final spacetimes. The semiclassical
approximation to the rate is then simply related to the
Euclidean action for the instanton.
 One such process, in
 which a cosmic string splits, with black holes appearing at the two ends,
 can be
 described approximately by a gravitational instanton based on the charged
 C-metric\foot{This has been previously noted by Gibbons \gibbons,
 though not the
 argument which follows below about how the gauge field behaves. }.

The Lorentzian
charged C-metric describes a pair of charged black holes accelerating away
from one another along a symmetry axis \kw, say the $z$-axis.  The C-metric
then has
conical singularities on the $z$-axis characterized by a deficit angle
$\delta_{in}$ on the inner part of the axis, between the two black holes,
and deficit angle
$\delta_{out}$ on the outer parts of the axis, extending from each black hole
out to $z=\pm\infty$.  These conical
singularities may be removed by introducing
a background magnetic field \ernst\ of the
appropriate strength to provide the force necessary to
accelerate the black holes.  The resulting metric, known as the Ernst metric,
has served as the starting point for calculations of the pair
creation rate for magnetically charged black holes in a background
magnetic field \gibbons,\gs,\dggh.

In this paper, however, we will work with the
C-metric directly, interpreting the conical singularities as a model for
a thin cosmic string along the $z$-axis.
It has recently been shown that
the conical singularity may  indeed be filled in with stress energy
corresponding to a real cosmic string \ruth.
For positive black hole mass, one has
$\delta_{in}<\delta_{out}$, implying that the
mass per unit length of the string
 is greater on the outer axis than on the inner axis.  The
corresponding difference in string tension between the inner and outer
axis
provides the force which accelerates the black holes.
The parameters of the C-metric may be chosen so that $\delta_{in}=0$,
corresponding to a string which breaks completely.\foot{The
(nonextreme) black holes which are produced have their horizons
identified to form
a wormhole in space. If $\mui = 0 $, the cosmic string
does not actually break, but simply passes through
the wormhole.}
More generally the string can `fray'.
In a real cosmic string, the magnetic flux is quantized.
If the string carries only a single unit
of flux, then it must `break' entirely.  If it carries multiple units of flux,
than it can fray by discrete amounts, corresponding to a given number of flux
quanta.

The Euclidean action for the C-metric is infinite, but the physical quantity
determining the rate of pair creation is the difference between this
action and that of an appropriate background geometry.
As for the Ernst instanton, we find that this difference
is given by a simple geometrical expression \hhr,
$\Delta I =-{1\over 4}(A_{bh} +\Delta A_{acc})$, where $A_{bh}$ is the
area of the black hole horizon and $\Delta A_{acc}$ is the area of the
acceleration horizon relative to the background.
For
small mass per unit length $\mu$ of the string,
the relative action determining the rate is given by
\eqn\rate{
\Delta I \simeq  {\pi m^2 \over \mu_{out}-\mu_{in} }
.}
The semiclassical approximation to the rate is then
$e^{-\Delta I}$.
\newsec{Cosmic Strings and Black Holes}

We begin by describing how a cosmic string can appear to end on a
black hole.
For definiteness, consider the Abelian-Higgs model coupled to gravity.  The
matter fields are a $U(1)$ gauge field $A_\mu$ and a charged scalar field
$\Phi$ with a Mexican hat potential.  The cosmic string is the familiar
Nielsen-Olsen vortex. In the simplest case,
one unit of magnetic flux runs along the center of the
vortex.  The scalar field far from the string is $\Phi\approx v\exp(i\phi)$,
where $v$ is the vev and $0\le\phi \le 2\pi$ is an angular coordinate around
the
string. So the phase of $\Phi$ has unit winding number going
around a large loop linking the string.

Now suppose the cosmic string enters a black hole.
On a constant time slice, the horizon is topologically a 2-sphere.
For simplicity, the natural thickness of the string will be taken much
smaller than the radius of the black hole.
The string pierces the horizon at some point $S$ (``south pole").
Take a loop on the horizon around $S$ much larger than the string
thickness but smaller than the black hole.  Around this loop, $\Phi$
winds once in phase. Deform the loop, and attempt to shrink it to the
antipodal point $N$ (``north pole").

It seems as if there will be trouble because of the winding number of
$\Phi$ in phase.  But phase is gauge dependent, and this winding
number can be unwound by a suitable gauge transformation
\eqn\gauge{
	\Phi' = U\Phi, \qquad
		e A_\mu'  = e A_\mu + iU^{-1}\partial_\mu U    }
which merely implies that we need a nontrivial $U(1)$ bundle.

To be explicit, take a slightly-larger-than-hemispherical
gauge patch on the event
horizon, about $S$.  Take a similar patch about $N$.  The two patches
are to overlap along a closed (``equatorial") strip.  To define a bundle
we give a gauge transformation $U$ on the overlap, to take us from the
$S$ patch to the $N$ patch;  a nontrivial bundle is defined by a
topologically nontrivial $U$.  To unwind the phase, it suffices to take
$U=\exp(-i\phi)$ where $0\le\phi  \le 2\pi$ is an angular coordinate
(``longitude") on the horizon that
runs around the strip.  The vector potential can be taken as $A_{\mu}=0$ in
the $N$ patch, which will gauge-transform in the overlap region into
the required vector potential in the $S$ patch.  This completes the
construction.

We have constructed here a field configuration topologically equivalent
to the Wu-Yang monopole \ryder.  In the Wu-Yang monopole the magnetic flux
is spread uniformly over the $2$-sphere, whereas here
the magnetic flux is all gathered up and
concentrated into a narrow flux tube at $S$.  The $U(1)$ bundle
we have constructed is precisely the well-known bundle that arises from
the Hopf fibration of the 3-sphere.

Now consider possible time dependence.
The magnetic flux crossing any closed two-surface is absolutely
conserved, according to a topological conservation law.  Thus
the flux entering each separate black hole is absolutely conserved, and
if a black hole terminates a string at one time, that black hole
must always terminate a string.  The only way of circumventing this
restriction in classical gravity is to allow the black holes
themselves to merge, with the total flux remaining conserved.
In quantum gravity, black holes themselves can
be created or destroyed in pairs, and the topological conservation law
simply constrains the total magnetic flux of both holes to be zero,
while individually the fluxes may be nonzero. Thus, through the creation by
quantum tunneling of a black hole pair along a cosmic string,
the string can break.

The same process can occur in a wide class of theories that admit local
cosmic strings, i.e., in which the vacuum manifold has a nontrivial
$\pi_1$.  Some such theories will also admit monopoles on which
cosmic strings can end, and in such theories cosmic strings can
also break through creation of monopole pairs \vilenkin .
However, string breaking by black hole pairs is often allowed,
even if the theory admits no such monopoles;  a sufficient
condition is that the unbroken symmetry group be connected\foot{To
see what can happen if the unbroken symmetry group is disconnected, e.g. $Z_n$,
consider a theory containing a Higgs field with
 charge $ne$, $n>1$, which condenses \qhair.
If its phase only wraps once
around the string, then the would-be gauge transformation
to unwrap it would be $U = exp(-i\phi/n)$.  However, this is not
single valued, and would, for instance, cause trouble with
other, singly charged fields.}
\jpav,\jppriv .

\newsec{Splitting Strings}
As discussed in the introduction, the instanton describing the pair creation of
black holes along a cosmic
string is given by the Euclidean C-metric. This metric
and gauge potential $V_\mu$ are given by
\eqn\cmetric{ \eqalign{
ds^2 &=r^2 \left ( -G(y) \({\beta\over 2\pi}\)^2 d\tau ^2
-{dy^2 \over G(y)} +\kappa ^2
G(x) d\phi ^2 +{dx^2 \over G(x)} \right ) \cr
V_{\phi}&=\kappa q (x-\xd) \cr
r&={1\over A(x-y)} ,\qquad G(x) = 1-x^2-2mAx^3-q^2A^2x^4\cr
}}
where $0\leq \phi \leq 2\pi $ and $0\leq \tau \leq 2\pi$.
The function $G(\xi)$ has four roots which we
shall label $\xa  \leq \xb <\xc <\xd  $. Hence the coordinate ranges are $\xb
\leq y
\leq \xc,\ \xc \leq x \leq \xd $. The black hole horizon is at $y=\xb$ and the
acceleration horizon is at $y=\xc $. The inner strut is at $x=\xd$ and the
outer
strut at $x=\xc$. Spatial infinity is at the point where $x=y=\xc$, and so
$r\rightarrow \infty$. The black holes carry magnetic charge $q$
under the unbroken $U(1)$ gauge field $V_\mu$.  The reader should note
that this gauge field is distinct from the broken gauge field $A_\mu$ from
which
the cosmic string is constructed.  The presence of this second gauge
field is required below in order to construct a smooth instanton.

In the model of a cosmic string by flat space minus a wedge,
the mass per unit length of the string is equal to $\delta /8\pi $, where
$\delta$ is the deficit angle. In terms of the metric coefficients, the deficit
angle
on the outer axis is given by
\eqn\defdo{ \do = 2\pi \(1- {\kappa \over 2}|G'(\xc )|\) }
and the deficit angle on the inner axis is
\eqn\defdi{\di =2\pi \(1- {\kappa \over 2} |G'(\xd )|\).}
Clearly, from the symmetrical form of the metric, there
are also nodes in the $\tau -y$ plane for generic choices of parameters.
Unlike the conical singularities along the axis, these nodes cannot be
interpreted as approximations to a smooth cosmic string. Instead, they
represent points where the field equations are no longer satisfied. In
the usual instanton approximation, one requires that the equations hold
everywhere, and so these singularities must be avoided. There are two ways
to achieve this. First, one can set
\eqn\smooth{ G'(\xb )= -G'(\xc ), \qquad \beta = 4\pi / G'(\xc ).}
This requires that $q=m$ in the definition of $G(x)$ and implies that
$ \xc -\xa = \xd-\xb.$
Geometrically, this corresponds to pair creating nonextreme black holes
with their horizons identified to form a wormhole \gs.
The surface gravities, or temperatures, of the black hole and
acceleration horizons are equal.
Alternatively, one can consider extremal black holes where $\xi_1 = \xi_2$
\dggh .
In this case, the black hole  horizon is infinitely far away. The conical
singularity on the acceleration horizon will be absent provided we again set
$\beta = 4\pi/ G'(\xc )$.

Consider a cosmic string of a given $\muo $, or equivalently, a given $\do$.
We want to compute the rate at which extreme and nonextreme black holes are
pair
produced with a string of deficit angle $\di <\do$ between them.
So we need to evaluate the Euclidean action for the C-metric
with these parameters. The metric \cmetric\ contains five parameters: $m, q, A,
\beta, \k$. Two of these are fixed by \smooth\ (or the analogous conditions
for extreme black holes).
Two are fixed by our choice of $\do$ and $\di$. The remaining parameter
can be thought of as the charge of the created black holes and remains
arbitrary.

The Euclidean action for the Einstein-Maxwell theory is given by
\eqn\action{
I = {1\over 16\pi}\int_M [-R+F^2] -{1\over 8\pi}\int_{\partial M}K
}
This is infinite for \cmetric, but the physically
meaningful quantity is the difference between the action for the
C-metric, and a reference background. The appropriate background here is
flat space minus a wedge with deficit angle $\do$. As discussed earlier,
we are viewing the conical singularity in the C-metric and the background as
an approximation to a thin smooth string composed of gauge and scalar fields,
which satisfy their field equations everywhere. Thus, in evaluating the
action, there is no need to introduce additional boundaries around the
conical singularity.
As discussed in \by, \teit, \hh, the action is conveniently evaluated on
a solution by rewriting it in
Hamiltonian form. The surfaces of constant $\tau$ intersect on the horizons,
and these points of intersection
must be treated separately. Evaluating the action
in a neighborhood of the horizon yields a contribution $-A/4$, where $A$ is
the horizon area, so
one obtains \hh
\eqn\hhaction{ \D I = \beta H - {1\over 4} \D A_{acc} - {1\over 4} A_{BH} }
where $H$ is the total energy of the C-metric relative to the background,
and $\D A_{acc}$ is the difference between the area of the acceleration
horizons
in the C-metric and the background.
$H$ is the sum of a term which is pure constraint and, hence, vanishes
on a solution, plus an extrinsic curvature boundary term given below.
For the extremal black hole of metric \cmetric ,
the horizon is infinitely far away, and so the
surfaces of constant $\tau$ do not intersect there. As as result, there
is no term ${1\over 4} A_{BH}$ in the action.

To evaluate the first two
terms in \hhaction, we need to match the C-metric and the background metric
on a large sphere near infinity. The sphere is defined by
$x-y =\epsilon$, and we are interested in the limit $\epsilon
\rightarrow 0$. As in \hhr, we change to new coordinates $\chi ,\epsilon$ with
$x=\xc +\epsilon\chi ,\  y=\xc +\epsilon (\chi -1)$ , where
$0\leq \chi \leq 1$. Then the induced
metric on the two surface $d\tau =d\epsilon =0$ is
\eqn\twometric{
{}^2 ds^2={1\over \epsilon A^2 \gp} \left[ 4\ko ^2  \chi\(1+ {\ep\over 2}
 {\gpp \over \gp }\chi \)d\phi ^2 - {d\chi ^2 \over \chi (\chi - 1)}
\right]
}
The background metric can be described by \cmetric\ with $m=q=0$.
We now require that the metric \twometric\ agree with the metric
induced on the surface $x-y =\bar{\ep}$
in the background where
 $\bar{G}(x) =1-x^2 $ and $\bar{\xc} =-1$.
This will be the case provided
\eqn\matchcond{\gp A^2\ep =2 \bar{A}^2 \bar{\ep},\ \ {\rm and} \ \
  -\ep {\gpp \over \gp }=
\bar{\ep} }
where $\bar{A}$ is the parameter appearing in the background metric.

On a solution,
the Hamiltonian in \hhaction\ is given by $H =
\int N ({ }^{(2)}K - { }^{(2)}\bar{K})$, where ${ }^{(2)}K$ is the extrinsic
curvature of the  boundary in the $\tau$ = constant surface.
The components of the normal to this surface are given by $n^x = -{G(x)\over
r\sqrt{ G(x) -G(y) }} \ , n^y =-{G(y)\over r\sqrt{G(x) -G(y) }}$ and one finds
\eqn\ktwo{
{ }^{(2)}K =D_i n^i = A\sqrt{\ep \gp} \left( 1+\ep
{\gpp\over \gp}(\chi - {3\over 4}) \right).
}
Subtracting the analogous expression for the extrinsic curvature in the
background,
and using the matching conditions \matchcond, one finds that
${ }^{(2)}K -{ }^{(2)}\bar{K} =0(\ep^2)$.
  From \twometric, we see that
$\sqrt{ { }^{(2)}g}$ goes like $\ep ^{-1}$. The lapse behaves like
$N = O(\ep^{-1/2})$. Therefore
the energy term in the action vanishes as $\ep \rightarrow 0$.

We now compute $\D A_{acc}$.
Since the area of each acceleration horizon is infinite,
we integrate out to the surface $x=\xc + \ep $, subtract,
and then take $\ep$ to zero:
\eqn\accarea{
A_{acc} =\int_0^{2\pi} \kappa d\phi \int _{\xc +\ep} ^{\xd} {dx \over A^2
(x-\xc )^2 }={2(2\pi - \do)
\over \ep A^2 \gp }\left( 1- {\ep\over \xd -\xc }\right)
}
Subtracting the similar expression for $\bar{A}_{acc}$ and using \matchcond\
gives
\eqn\accareadiff{
\Delta A_{acc} = A_{acc} -\bar{A}_{acc}
%
= -{2(2\pi - \do) \over A^2 \gp} \left( {1\over \xc -\xa } + {1\over \xc-\xb }
\right)
}
The area of the black hole horizon is
\eqn\bharea{
A_{BH}={2(2\pi - \do)\over A^2 \gp}
\( {1\over \xi_3 - \xi_2} - {1\over \xi_4 -\xi_2} \) }
Combining these  and using \smooth , gives the total physical action,
\eqn\Iareas{
\Delta I
 = {2\pi - \do  \over A^2 \gp  (\xi_3 - \xi_1)} }
This formula is also valid for
the extremal instanton since in this case
$\Delta I = -{1\over 4} \Delta A_{acc}$ and $\xi_2 = \xi_1$.

For small $mA$ we can find a simple expression for this action.
If we fix $\do$, the deficit angle of the string at infinity, and $\di$, the
deficit angle of the
string connecting the black holes, then we can expand $G'(\xc)$
and $G'(\xd)$ to first order in $mA$ and use \defdo\ and \defdi\
to solve for $mA$.
The result is
\eqn\newlaw{ mA = {1\over 8\pi} (\do - \di) = \muo - \mui}
This says that the black holes satisfy Newton's law. The acceleration is
determined by the net tension in the strings connecting the black holes.
Expanding the terms in the action \Iareas\
in powers of $mA$ and using this result
we obtain
\eqn\smallaction{
\Delta I\simeq  {\pi m^2 \over \muo -\mui }
}
The rate, $e^{-\Delta I}$, is largest for the string breaking $\mui =0$.
This makes sense because
roughly the mass of the black holes must come from the missing mass  of the
string, so $\mui =0$
corresponds to the black holes tunneling out at the smallest separation,
which one
expects
for a quantum event.  The rate increases for a more massive external string,
and
the rate vanishes when $\muo =\mui$, which says that one cannot pair create
black holes without taking some energy away from the cosmic string.

\newsec{Real Strings}

The process we have discussed could have cosmological
significance.  It is well known that any process that turns cosmic strings
into black holes (or other massive remnants) might seriously disrupt
cosmic string
cosmology.  Note that black holes are always left behind;  in a closed
loop of strings, a nucleated black hole pair will race around the string,
consume it entirely, and collide to leave behind one (or perhaps more)
black holes.  If multiple nucleations happen, multiple collisions will
occur.

One can, however, substitute numbers corresponding to grand unified strings
into \smallaction\ and find that the rate for breaking cosmic strings by
this mechanism is far too small to be of cosmological significance.
For a Higgs vacuum expectation
value $v\sim 10^{16}$GeV and self coupling $\lambda \sim 1$, we must take
the black hole to have mass $m>> 10^3\ m_{pl}$ in order for the thin
string limit implicit in the use of the C-metric to be valid.
This implies $\mu \sim v^2 \sim 10^{-6} m_{pl}^2.$
We then have $\Delta I >> 10^{12}$, yielding
an infinitesimally small rate.

However, this estimate of the rate is only a lower limit.
The most likely tunneling event actually falls outside the class described by
the C-metric.  This would be to pair create the smallest possible black holes
which can swallow the flux from the string.  One can estimate the size of such
a
black hole as having mass equal to a single quanta of magnetic charge,
making it extremal.
For the parameters assumed above, such a black hole would be small
on the scale of a flux tube, so we
would need another method for estimating the rate of production.

It is interesting to speculate about the production rate for
black holes with mass not equal to charge. For a general choice of
$q$ and $m$ in \cmetric, there is a nodal singularity at the Euclidean black
hole horizon. However, this singularity is
integrable--it is only a two dimensional delta-function in the
curvature. Evaluating the action \action\ in the neighborhood of a
horizon, one still finds that the contribution is ${1\over 4}A_H$,
using the Gauss-Bonnet theorem \teit. Therefore, the action evaluated
on any of the C-metrics is given by the basic formulae \hhaction .
Further, combining \accareadiff\ and \bharea , one finds that
for any of the C-metrics {\it except} the extremal black hole
case, the action is given by
\eqn\lastI{
\Delta I  = {(2\pi -\do)\over 2A^2 G'(\xc )}
\left( {1\over \xc -\xa } +{1\over \xd -\xb }\right)
}
Finally, one finds that the value of $\Delta I$ for small $mA$
given in \smallaction\
is the same for all values of
$q,m$.

These nonsmooth C-metrics are not solutions everywhere, and
so they do not have the usual instanton interpretation. However,
since they fail to be a solution in a very mild way, and the
``answer" they give for the rates is of exactly the same form
as the smooth case, it is tempting to speculate that they
do give the leading contribution to the pair production rate
for general $q,m$. This is an issue for further consideration.

NOTE ADDED: After this work was completed, two papers appeared which
discuss black hole pair creation and cosmic strings. The first \hr\
considers the case $\mui=0$, and asserts that the calculation does
not apply to topologically stable strings. We clearly disagree with this
statement. The second \emp\ adds a background magnetic field  and sets $\muo=0$
(leaving $\mui \ne 0 $), but does not discuss the applicability to real cosmic
strings.

\bigskip
\centerline{Acknowledgments}
We wish to thank Fay Dowker, Jerome Gauntlett and Ruth Gregory
for helpful discussions, and John Preskill for pointing out an incorrect
assertion in an earlier version.
This work was supported in part by NSF grants
PHY-9008502, THY-8714-684-A01
and PHY-9407194. DME is grateful for the hospitality
of the Aspen Center for Physics.
DK and JT thank the Institute for Theoretical Physics and the UCSB
physics department for their hospitality.

\listrefs
\end